\documentclass[aps,prc,superscriptaddress,nofootinbib,showpacs,floatfix,longbibliography]{revtex4}
\usepackage{url}
\usepackage{cancel}
\usepackage[colorlinks,linkcolor=blue,citecolor=blue,filecolor=black,urlcolor=blue]{hyperref}
\usepackage{epsfig,graphics}
\usepackage{bm}
\usepackage{amssymb}
\usepackage{amsmath}
\usepackage{multirow}
\usepackage{float}
\usepackage{harpoon}
\usepackage{MnSymbol}
\usepackage{appendix}
\usepackage{color}
\usepackage{hyperref}

\newcommand{\nch} {N_{\mathrm{ch}}}

\newcommand{\pT} {p_{\mathrm{T}}}
\newcommand{\lr}[1]{\left\langle #1\right\rangle}

\newcommand{\npart}{N_{\mathrm{part}}}

\begin{document}
\title{Impact of nuclear shape fluctuations in high-energy heavy ion collisions}
\newcommand{\bnl}{Physics Department, Brookhaven National Laboratory, Upton, NY 11976, USA}
\newcommand{\sbu}{Department of Chemistry, Stony Brook University, Stony Brook, NY 11794, USA}
\author{Aman Dimri}\email{aman.dimri@stonybrook.edu}\affiliation{\sbu}
\author{Somadutta Bhatta}\affiliation{\sbu}
\author{Jiangyong Jia}\email{jiangyong.jia@stonybrook.edu}\affiliation{\sbu}\affiliation{\bnl}
\date{\today}
 
\begin{abstract}
The shape of atomic nuclei is often interpreted to possess a quadrupole deformation that fluctuates around some average profile. We investigate the impact of nuclear shape fluctuations on the initial state geometry in heavy ion collisions, particularly its eccentricity $\varepsilon_2$ and inverse size $d_{\perp}$, which can be related to the elliptic and radial flow in the final state. The fluctuation in overall quadrupole deformation enhances the variances and modifies the skewness and kurtosis of the $\varepsilon_2$ and $d_{\perp}$ in a controllable manner. The fluctuation in triaxiality reduces the difference between prolate and oblate shape for any observable, whose values, in the large fluctuation limit, approach those obtained in collisions of rigid triaxial nuclei. The method to disentangle the mean and variance of the quadrupole deformation is discussed.
\end{abstract}

\maketitle
\section{Introduction}
Ultra-relativistic heavy ion physics aims to understand the dynamics and properties of the Quark-Gluon Plasma (QGP) created in collisions of atomic nuclei at very high energy~\cite{Busza:2018rrf}. Achieving this goal is currently limited by the lack of understanding of the initial condition, i.e. how the energy is deposited in the overlap region before the formation of the QGP~\cite{Bernhard:2016tnd}. The energy deposition process is not calculable from first principles and is often parameterized via phenomenological approaches with multiple free parameters~\cite{Giacalone:2022hnz}. On the other hand, heavy atomic nuclei are well-studied objects interpreted to exhibit a wide range of shapes and radial profiles~\cite{ring,Moller:2015fba,Scamps:2020fyu}, which are often characterized by a few collective nuclear structure parameters such as average quadrupole and octupole deformations, nuclear radius, and skin thickness. One can leverage species with similar mass numbers but different structures, such as isobars, to directly probe the energy deposition mechanism and hence constrain the initial condition. The efficacy of this approach has been investigated recently~\cite{Jia:2021oyt,Nijs:2021kvn,Jia:2022qgl}.

One good example demonstrating this possibility is the $^{96}$Ru+$^{96}$Ru and $^{96}$Zr+$^{96}$Zr collisions, recently carried out by the STAR Collaboration at the relativistic heavy ion collider (RHIC)~\cite{STAR:2021mii,chunjianhaojie}. Ratios of many bulk observables between the isobars, such as harmonic flow $v_n$, charged particle multiplicity $\nch$, and average transverse momentum $\lr{\pT}$, have been measured, which show significant and observable- and centrality-dependent deviation from unity. Model studies show that these ratios are insensitive to final-state effects and are controlled mainly by the differences of the collective nuclear structure parameters between $^{96}$Ru and $^{96}$Zr~\cite{Zhang:2022fou}. Comparing calculations with experimental data, Refs.~\cite{Zhang:2021kxj,Jia:2021oyt} have estimated structure parameters that are broadly consistent with general knowledge from low energy side. However, these studies also suggest a sizable octupole collectivity for Zr, not predicted by mean field structure models~\cite{Cao:2020rgr}. The rich and versatile information from isobar or isobar-like collisions provides a new constraint on the heavy ion initial condition and a new way to probe nuclear structure at high energy~\cite{Bally:2022vgo}.

However, it is important to point out that atomic nuclei in the ground state often do not have a static shape, but can fluctuate around an average profile. The potential energy surface of such species usually has shallow minima as a function of deformation parameters, such as the triaxial quadruple deformations $\beta$ and $\gamma$. The ground state nuclear wave function is often treated as a mixture of configurations with different $(\beta,\gamma)$ values~\cite{RevModPhys.75.121,Bender:2008zv,Rodriguez:2010by}. Then there is the phenomena of shape coexistence, which happens when the same nuclei can have multiple low-lying states with widely different shapes but small energy differences~\cite{RevModPhys.83.1467}. From the nuclear structure side, the quadrupole fluctuations can be estimated from the sum rules of matrix elements of various moments of quadrupole operators that can be measured experimentally~\cite{Kumar:1972zza,Poves:2019byh}. From the heavy ion collision side, the shape fluctuations can be accessed using multi-particle correlations, which probe moments of the nucleon position in the initial condition~\cite{Jia:2021tzt}. For instance, the elliptic flow $v_2$ in each event is approximately proportional to the elliptic eccentricity $\varepsilon_2$, $v_2\approx k \varepsilon_2$, calculable from participating nucleons~\cite{Teaney:2010vd}. Therefore, the fluctuations of flow are related to fluctuations of quadruple deformation via their respective moments: $\lr{v_2^{m}}\approx k^m \lr{\varepsilon_2^m}\propto  \lr{\beta^m}, m=2,4..$. In principle, one could constrain the mean and variance of quadrupole fluctuations from $\lr{\beta^2}$ and $\lr{\beta^4}$, which in turn can be determined from $\lr{v_2^{2}}$ and $\lr{v_2^{4}}$. An early study investigated the role of nuclear shape vibration in near-spherical $^{208}$Pb, which was found to have a significant impact on $\lr{\varepsilon_2^2}$ in central collisions~\cite{Zakharov:2020irp,Zakharov:2021lux}.

This paper extends our previous study~\cite{Jia:2021qyu} to investigate the influence of fluctuations of quadruple deformation parameters $(\beta, \gamma)$ to several selected two-, three- and four-particle heavy-ion observables. We first derive simple analytical relations between these observables and the means and variances of $(\beta,\gamma)$. We then perform a more realistic Glauber model simulation, assuming Gaussian fluctuations, to quantify the region of validity of these relations. We discuss the sensitivity of these observables on the nuclear shape, as well as the prospect of separating the average shape from shape fluctuations.  

\section{Expectation and model setup}
We consider the eccentricity vector ${\bm \epsilon_2}\equiv\varepsilon_2 e^{2i\Phi_2}$ and inverse transverse size $d_{\perp}$, which are estimators for elliptic flow $V_2\equiv v_2 e^{2i\Psi_2}$ and average transverse momentum $\lr{\pT}$ or radial flow, calculated from the transverse position of nucleon participants in each event,
\begin{align}\label{eq:1}
{\bf \epsilon_2} = - \frac{\lr{r_{\perp}^2 e^{i2\phi}}}{\lr{r_{\perp}^2}}\;, d_{\perp} =\sqrt{\npart/\lr{r_{\perp}^2}},
\end{align}
where $r_{\perp}$ is the transverse radius and $\npart$ is the number of participating nucleons. Following the heuristic argument from Ref.~\cite{Jia:2021qyu}, for collisions of nuclei with small quadrupole deformation, the eccentricity vector and $d_{\perp}$ in a given event have the following leading-order form:
\begin{align}\nonumber
\frac{\delta d_{\perp}}{d_{\perp}} &\approx  \delta_d + p_0(\Omega_p,\gamma_p)\beta_p+p_0(\Omega_t,\gamma_t)\beta_t\;,\\\label{eq:2}
{\bm \epsilon}_2 &\approx {\bm \epsilon}_0 + {\bm p}_{2}(\Omega_p,\gamma_p)\beta_p +{\bm p}_{2}(\Omega_t,\gamma_t)\beta_t,
\end{align}
where the scalar $\delta_d$ and vector ${\bm \epsilon}_0\equiv\varepsilon_0 e^{2i\Phi_{2;0}}$ are values for spherical nuclei. Here, we are considering the general situation where the projectile and target, denoted by subscripts ``$p$'' and ``$t$'', have different deformation values. In Eq.~\eqref{eq:2}, $p_0$ and ${\bm p}_{2}$ are phase space factors, which depend on $\gamma$ and the Euler angles $\Omega$. 

Since the fluctuations of $\delta_d$ (${\bm \epsilon}_0$) are uncorrelated with $p_0$ (${\bm p}_{2}$), an average over collisions with different Euler angles is expected to give the following leading-order expressions for the variances, skewness, and kurtosis of the fluctuations
\begin{subequations}\label{eq:3}
\begin{align}
c_{\mathrm{d}}\{2\}&\equiv\lr{\left(\frac{\delta d_{\perp}}{d_{\perp}}\right)^2} = \lr{\delta_d^2} + \lr{p_0(\gamma_p)^2}\beta_p^2+\lr{p_0(\gamma_t)^2}\beta_t^2\;,\\
c_{2,\epsilon}\{2\}&\equiv\lr{\varepsilon_2^2} = \lr{\varepsilon_0^2} +  \lr{{\bm p}_{2}(\gamma_p){\bm p}_{2}^*(\gamma_p)}\beta_p^2+  \lr{{\bm p}_{2}(\gamma_t){\bm p}_{2}^*(\gamma_t)}\beta_t^2\;,\\
\mbox{Cov}&\equiv\lr{\varepsilon_2^2 \frac{\delta d_{\perp}}{d_{\perp}}} = \lr{\varepsilon_0^2\delta_d} +\lr{p_0(\gamma_p){\bm p}_{2}(\gamma_p){\bm p}_{2}(\gamma_p)^*}\beta_p^3+\lr{p_0(\gamma_t){\bm p}_{2}(\gamma_t){\bm p}_{2}(\gamma_t)^*}\beta_t^3\;,\\
c_{\mathrm{d}}\{3\}&\equiv \lr{\left(\frac{\delta d_{\perp}}{d_{\perp}}\right)^3} = \lr{\delta_d^3} +\lr{p_0(\gamma_p)^3}\beta_p^3+\lr{p_0(\gamma_t)^3}\beta_t^3\;,\\
c_{2,\epsilon}\{4\}&\equiv \lr{\varepsilon_2^4}-2\lr{\varepsilon_2^2}^2  = \lr{\varepsilon_0^4}-2\lr{\varepsilon_0^2}^2 +\left(\lr{{\bm p}_{2}^2{\bm p}_{2}^{2*}}\lr{\beta^4}-2\lr{{\bm p}_{2}{\bm p}_{2}^{*}}^2\lr{\beta^2}^2\right)_p+\left(\lr{{\bm p}_{2}^2{\bm p}_{2}^{2*}}\lr{\beta^4}-2\lr{{\bm p}_{2}{\bm p}_{2}^{*}}^2\lr{\beta^2}^2\right)_t\;.
\end{align}
\end{subequations}
Note that the ``$\lr{}$'' in the above equation is taken over the Euler angles, and these quantities correspond to the results for fixed deformation values.

Previous studies have demonstrated that the moments $\lr{p_0^2}$, $\lr{{\bm p}_{2}{\bm p}_{2}^*}$, and $\lr{{\bm p}_{2}^2{\bm p}_{2}^{2*}}$ are independent of $\gamma$, while $\lr{p_0{\bm p}_{2}{\bm p}_{2}^*}$ and $\lr{p_0^3}$ have an leading order dependence, $c+b\cos(3\gamma)$. Here, $c\ll b$ for $\lr{p_0{\bm p}_{2}{\bm p}_{2}^*}$, whereas $c\lesssim b$ for $\lr{p_0^3}$~\cite{Jia:2021qyu}. In the presence of quadrupole fluctuations, we further need to average these quantities over ``independent'' fluctuations for projectile and target. Assuming that the fluctuations of the projectile and target are sampled from the same probability density distributions, we have
\begin{subequations}\label{eq:4}
\begin{align}
\lr{\left(\frac{\delta d_{\perp}}{d_{\perp}}\right)^2} &= a_0+ \frac{b_0}{2} \left(\lr{\beta_p^2}+\lr{\beta_t^2}\right) = a_0+ b_0\lr{\beta^2}\;,\\
\lr{\varepsilon_2^2} &= a_1+\frac{b_1}{2}(\lr{\beta_p^2}+\lr{\beta_t^2})=a_1+b_1\lr{\beta^2}\;,\\
\lr{\varepsilon_2^2 \frac{\delta d_{\perp}}{d_{\perp}}}&=a_2-\frac{1}{2}\left(\lr{(c_2+b_2\cos(3\gamma_p))\beta_p^3}+\lr{(c_2+b_2\cos(3\gamma_t))\beta_t^3}\right)=a_2-\lr{(c_2+b_2\cos(3\gamma))\beta^3}\;,\\
\lr{\left(\frac{\delta d_{\perp}}{d_{\perp}}\right)^3} &= a_3+ \frac{1}{2} \left(\lr{(c_3+b_3\cos(3\gamma_p))\beta_p^3}+\lr{(c_3+b_3\cos(3\gamma_t)\beta_t^3}\right)=a_3+ \lr{(c_3+b_3\cos(3\gamma))\beta^3}\;,\\
\lr{\varepsilon_2^4}-2\lr{\varepsilon_2^2}^2 &= a_4+\frac{b_4}{2}\left(\lr{\beta_p^4}+\lr{\beta_t^4}\right)-\frac{c_4}{2}\left(\lr{\beta_p^2}^2+\lr{\beta_t^2}^2\right)=a_4+b_4\lr{\beta^4}-c_4\lr{\beta^2}^2\;,
\end{align}
\end{subequations}
where the averages are performed over fluctuations in $\beta$ and $\gamma$, and the coefficients $a_n$, $b_n$ and $c_n$ are centrality-dependent positive quantities satisfying $c_2\ll b_2$ and $c_3\lesssim b_3$~\cite{Jia:2021qyu}. The quantities in Eq.~\eqref{eq:4} can be relate directly to the final state observables, $\lr{v_2^2}$, $\lr{(\delta \pT/\lr{\pT})^2}$, $\lr{v_2^2\frac{\delta \pT}{\lr{\pT}}}$, $\lr{(\delta \pT/\lr{\pT})^3}$ and $\lr{v_2^4}-2\lr{v_2^2}^2$, respectively.

A crude numerical estimation can be obtained in the liquid-drop model, where the nucleon density distribution has a sharp surface. For head-on collisions with zero impact parameter, it predicts the following simple relations~\cite{Jia:2021qyu},
\begin{align}\nonumber
\frac{\delta d_{\perp}}{d_{\perp}}&=\sqrt{\frac{5}{16 \pi}} \beta_{2}\left(\cos(\gamma) D_{0,0}^{2}(\Omega)+\frac{\sin(\gamma)}{\sqrt{2}}\left[D_{0,2}^{2}(\Omega)+D_{0,-2}^{2}(\Omega)\right]\right)\;,\\\label{eq:5}
\;{\bm \epsilon}_{2}&=-\sqrt{\frac{15}{2 \pi}} \beta_{2}\left(\cos(\gamma) D_{2,0}^{2}(\Omega)+\frac{\sin(\gamma)}{\sqrt{2}}\left[D_{2,2}^{2}(\Omega)+D_{2,-2}^{2}(\Omega)\right]\right)\;,
\end{align}
where the $D^{l}_{m,m'}(\Omega)$ are the Wigner matrices. The analytical results obtained for various cumulants are listed in Table~\ref{tab:1}. They provide approximate estimates for the values of $b_n$ in most central collisions ($c_n=0$ in the liquid-drop model).

\begin{table}[!h]
\centering
\begin{tabular}{c||c}\hline
Cumulants& Liquid-drop model estimate\\\hline
\multirow{2}{*}{$\lr{(\delta d_{\perp}/d_{\perp})^2}$} &  \multirow{2}{*}{$\frac{1}{32\pi}\lr{\beta^2}$}\\ &\\\hline
\multirow{2}{*}{$\lr{(\delta d_{\perp}/d_{\perp})^3}$} & \multirow{2}{*}{$\frac{\sqrt{5}}{896 \pi^{3/2}}\lr{\cos(3\gamma)\beta^3}$} \\ &\\\hline
\multirow{2}{*}{$\lr{(\delta d_{\perp}/d_{\perp})^4}-3\lr{(\delta d_{\perp}/d_{\perp})^2}^2$} & \multirow{2}{*}{$-\frac{3}{14336 \pi^{2}}\left(7\lr{\beta^2}^2-5\lr{\beta^4}\right)$}\\ &\\\hline
\multirow{2}{*}{$\lr{\varepsilon_2^2}$} & \multirow{2}{*}{$\frac{3}{4\pi}\lr{\beta^2}$}\\ &\\\hline
\multirow{2}{*}{$\lr{\varepsilon_2^4}-2\lr{\varepsilon_2^2}^2$} &  \multirow{2}{*}{$-\frac{9}{112\pi^2}\left(7\lr{\beta^2}^2-5\lr{\beta^4}\right)$}\\ &\\\hline
\multirow{2}{*}{$\left(\lr{\varepsilon_2^6}-9\lr{\varepsilon_2^4}\lr{\varepsilon_2^2}+12\lr{\varepsilon_2^2}^3\right)/4$}& \multirow{2}{*}{$\frac{81}{256\pi^3}\left[\lr{\beta^2}^3-\frac{45}{14}\lr{\beta^4}\lr{\beta^2} -\frac{1175}{6006}\lr{\beta^6}+\frac{25}{3003}\lr{\cos(6\gamma)\beta^6}\right]$}\\ &\\\hline
\multirow{2}{*}{$\lr{\varepsilon_2^2(\delta d_{\perp}/d_{\perp})}$} &  \multirow{2}{*}{$-\frac{3 \sqrt{5}}{112\pi^{3/2}} \lr{\cos(3\gamma)\beta^3}$}\\&\\\hline
\multirow{2}{*}{$\lr{\varepsilon_2^2(\delta d_{\perp}/d_{\perp})^2}-\lr{\varepsilon_2^2}\lr{(\delta d_{\perp}/d_{\perp})^2}$} & \multirow{2}{*}{$-\frac{3}{1792\pi^2}\left(7\lr{\beta^2}^2-5\lr{\beta^4}\right)$}\\&\\\hline
\multirow{2}{*}{$\lr{ {\bm \epsilon}_2^2{\bm \epsilon}_4^*}$} &\multirow{2}{*}{$\frac{45}{56\pi^2}\lr{\beta^4}$}\\&\\\hline
\end{tabular}
\caption{\label{tab:1} The leading-order results of various cumulants calculated for the nucleus with a sharp surface via Eq.~\eqref{eq:5}. The two nuclei are placed with zero impact parameter and results are obtained by averaging over random orientations.}
\end{table}

To make further progress, we consider the case where the fluctuations of $\beta$ and $\gamma$ are independent of each other. The observables in Eq.~\eqref{eq:4} and Table~\ref{tab:1} can be expressed in terms of central moments. Assuming Gaussian fluctuations with means ($\bar\beta$, $\bar\gamma$) and standard deviations ($\sigma_{\beta}$, $\sigma_{\gamma}$),  Eq.~\eqref{eq:4} becomes
\begin{subequations}\label{eq:6}
\begin{align}
\lr{\left(\frac{\delta d_{\perp}}{d_{\perp}}\right)^2} &=a_0+ b_0(\bar{\beta}^2+\sigma_{\beta}^2)\;,\\
\lr{\varepsilon_2^2} &=a_1+b_1(\bar{\beta}^2+\sigma_{\beta}^2)\;,\\
\lr{\varepsilon_2^2 \frac{\delta d_{\perp}}{d_{\perp}}}&=a_2-(b_2e^{-\frac{9 \sigma_{\gamma}^{2}}{2}}\cos (3\bar{\gamma})+c_2)\bar{\beta}(\bar{\beta}^2+3\sigma_{\beta}^2)\;,\\
\lr{\left(\frac{\delta d_{\perp}}{d_{\perp}}\right)^3} &= a_3+ (b_3e^{-\frac{9 \sigma_{\gamma}^{2}}{2}}\cos (3\bar{\gamma})+c_3)\bar{\beta}(\bar{\beta}^2+3\sigma_{\beta}^2)\;,\\
\lr{\varepsilon_2^4}-2\lr{\varepsilon_2^2}^2 &=a_4+b_4(\bar{\beta}^4+6\bar{\beta}^2\sigma_{\beta}^2+3\sigma_{\beta}^4)-c_4(\bar{\beta}^2+\sigma_{\beta}^2)^2\;,
\end{align}
\end{subequations}
where we have used the well-known expression for Gaussian smearing of an exponential function, $\lr{e^{in\gamma}} = e^{-\frac{n^2 \sigma_{\gamma}^{2}}{2}} e^{in\bar{\gamma}}$.

If the fluctuations of $\beta$ and $\gamma$ are non-Gaussian, one should also consider the higher cumulants of $\beta$. For example, $\lr{\beta^3} = \bar{\beta}(\bar{\beta}^2+3\sigma_{\beta}^2) + k_{3,\beta}$ and $\lr{\beta^4} = \bar{\beta}^4+6\bar{\beta}^2\sigma_{\beta}^2+3\sigma_{\beta}^4 + 4\bar{\beta}k_{3,\beta} +k_{4,\beta}$, where $k_{3,\beta} = \lr{(\beta-\bar{\beta})^3}$ and $k_{4,\beta}= \lr{(\beta-\bar{\beta})^4}-3\lr{(\beta-\bar{\beta})^2}^2$ are the skewness and kurtosis of the $\beta$ fluctuation. The expectation value of $\cos(n\gamma)$ can be expressed via the cumulant generating function of $\gamma$. Keeping the cumulants $k_{m,\gamma}$ up to leading order correction in skewness and kurtosis, $k_{3,\gamma}= \lr{(\gamma-\bar{\gamma})^3}$ and  $k_{4,\gamma}= \lr{(\gamma-\bar{\gamma})^4}-3\lr{(\gamma-\bar{\gamma})^2}^2$, we have,
\begin{align}\nonumber
\lr{\cos(n\gamma)} &= \frac{1}{2}\left(\lr{e^{in\bar{\gamma}}}+ \lr{e^{-in\bar{\gamma}}}\right)= \frac{1}{2}\left(\exp\left(\sum_{m=1}^{\infty} \kappa_{m,\gamma} \frac{(i n)^m}{m!}\right)+ \exp\left(\sum_{m=1}^{\infty} \kappa_{m,\gamma} \frac{(-i n)^m}{m!}\right)\right)\\\nonumber
&=\exp\left(\sum_{m=1}^{\infty} \kappa_{2m,\gamma} \frac{(-1)^m(n)^{2m}}{2m!}\right)\left[\cos\left(\sum_{m=1}^{\infty} \kappa_{2m+1,\gamma} \frac{(-1)^m(n)^{2m+1}}{(2m+1)!} +n\bar{\gamma}\right)\right]\\\label{eq:7}
&\approx  e^{-\frac{n^2 \sigma_{\gamma}^{2}}{2}+\frac{n^4 k_{4,\gamma}}{24}}\cos \left(n\bar{\gamma}-\frac{n^3}{6} k_{3,\gamma}\right)\approx  e^{-\frac{n^2 \sigma_{\gamma}^{2}}{2}}\left[\cos (n\bar{\gamma}) + \sin (n\bar{\gamma}) \frac{n^3}{6} k_{3,\gamma}\right](1+\frac{n^4}{24} k_{4,\gamma}).
\end{align}
Clearly, the net effect of skewness is a rotation of $\bar{\gamma}$ by $k_{3,\gamma}n^2/6$, while the net effect of kurtosis is to increase or decrease the overall variation with $\sigma_\gamma$ depending on its sign.

For a more realistic estimation of the influences of shape fluctuations, we perform a Monte-Carlo Glauber model simulation of $^{238}$U+$^{238}$U collisions. The setup of the model and the data used in this analysis are the same as those used in our previous work~\cite{Jia:2021tzt}. We simulate ultra-central collisions with zero impact parameter, where the impact of nuclear deformation reaches maximum. The nucleon distribution is described by a deformed Woods-Saxon function
\begin{align}\label{eq:8}
\rho(r,\theta,\phi)=\frac{\rho_0}{1+e^{\left[r-R(\theta,\phi)/a\right]}},\;R(\theta,\phi) = R_0\left(1+\beta [\cos(\gamma) Y_{2,0}(\theta,\phi)+ \sin(\gamma) Y_{2,2}(\theta,\phi)]\right),
\end{align}
where the nuclear surface $R(\theta,\phi)$ is expanded into spherical harmonics $Y_{2,m}$ in the intrinsic frame. Each nucleus is assigned a random $(\beta, \gamma)$ value, sampled from Gaussian distributions with means $(\bar{\beta},\bar{\gamma})$ and standard deviations $(\sigma_{\beta},\sigma_{\gamma})$. The nucleus is then rotated by random Euler angles before they are set on a straight line trajectory towards each other along the $z$ direction. Furthermore, three quark constituents are generated for each nucleon according to the quark Glauber model from Ref.~\cite{Loizides:2016djv}.  From this, the nucleons or the constituent quarks in the overlap region are identified, which are used to calculate $\varepsilon_2$ and $d_{\perp}$ defined in Eqs.~\eqref{eq:1}, and the results are presented as a function of deformation parameters.

For the study of the $\beta$ fluctuation, we fix $\gamma=0$ (prolate nucleus) and choose 11 values each for $\bar{\beta}^2$ and $\sigma_{\beta}^2$ from {0, 0.01,...,0.09, 0.1}. So a total of $11\times11=121$ simulations have been performed. For the study of the $\gamma$ fluctuation, we fix $\beta=0.28$ (the value for $^{238}$U) and choose seven $\bar{\gamma}$ and seven $\sigma_{\gamma}$ values: $\cos(3\bar{\gamma})=1,0.87,0.5,0,-0.5,0.87,-1$ and $\sigma_{\gamma}=0,\pi/18,2\pi/18,...,6\pi/18$, so a total of $7\times7=49$ simulation have been performed. For each case, about 50 Million events were generated and all the observables were calculated.  Our discussion is mainly based on the nucleon Glauber model, and the results from the quark Glauber model are included in the Appendix.

\section{Impact of triaxiality fluctuation}

Due to the three-fold symmetry of nuclear shape in triaxiality, the $\gamma$ dependence of a given observable can be generally expressed as $a_0+\sum_{n=1}^{\infty} \left[a_n \cos(3n\bar{\gamma})+b_n \sin(3n\bar{\gamma})\right]e^{-\frac{n^2 \sigma_{\gamma}^{2}}{2}}$. We further impose the condition that a random fluctuation for a triaxial nucleus does not impact the value of the observable, which is found to be true in our analysis, the $\gamma$ dependence becomes $a_0+\sum_{n=1}^{\infty} \left[a_n (\cos(3n\bar{\gamma})-\cos(3n\frac{\pi}{6}))+b_n(\sin(3n\bar{\gamma})-\sin(3n\frac{\pi}{6}))\right]e^{-\frac{n^2 \sigma_{\gamma}^{2}}{2}}$.

\begin{figure}[!h]
\begin{center}
\includegraphics[width=0.8\linewidth]{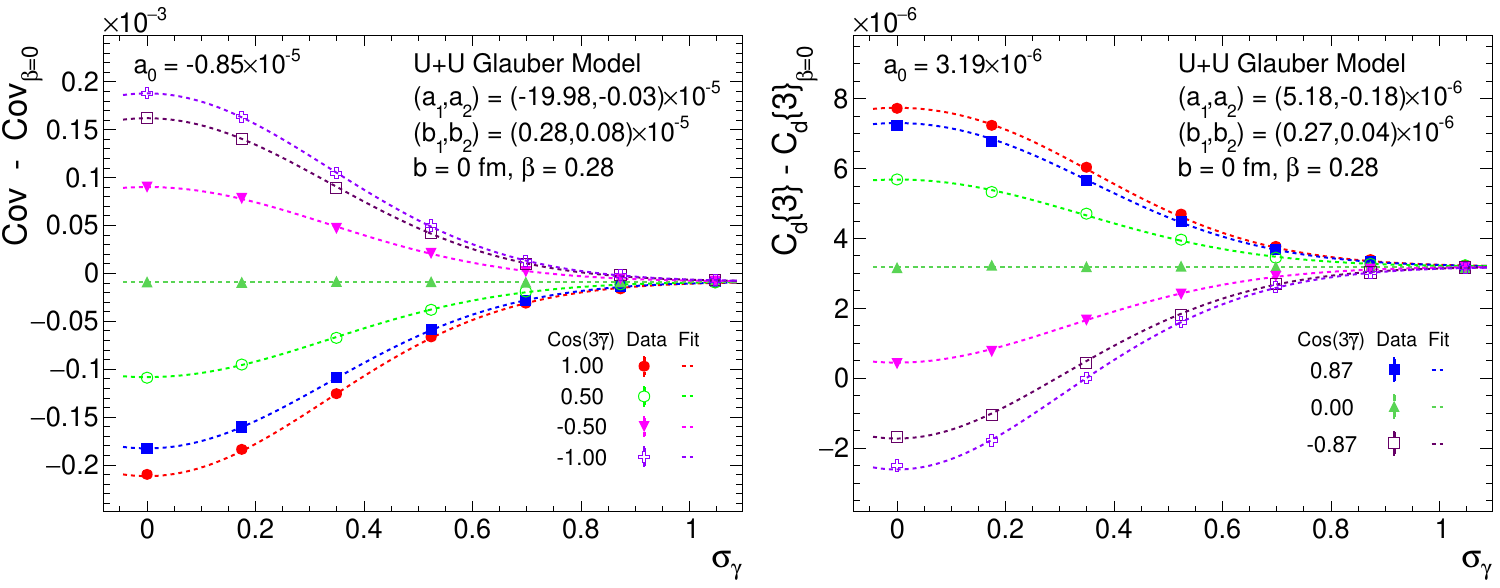}
\end{center}
\caption{\label{fig:1} 
The dependence of $\lr{\varepsilon_2^2 \frac{\delta d_{\perp}}{d_{\perp}}}$ (left) and $\lr{\left(\delta d_{\perp}/d_{\perp}\right)^3}$ (right) on smearing in triaxiality $\sigma_{\gamma}$ for different values of $\bar{\gamma}$. The lines indicate a simultaneous fit to Eq.~\eqref{eq:11} with the parameter values displayed on the plot.}
\end{figure}

We first discuss the impact of triaxiality fluctuation on three-particle observables $\lr{\varepsilon_2^2 \frac{\delta d_{\perp}}{d_{\perp}}}$ and $\lr{\left(\delta d_{\perp}/d_{\perp}\right)^3}$. We first subtract them by the values for the undeformed case\footnote{Random fluctuations of nucleon position in Glauber model naturally induce a small quadrupole deformation~\cite{Zakharov:2020irp}, which is subtracted and does not impact our discussion.}, to isolate the second term in Eq.~\eqref{eq:4} containing the triaxiality. Figure~\ref{fig:1} show the results obtained for different values of $\cos(3\bar{\gamma})$ as a function of $\sigma_{\gamma}$. The values for a pure triaxial nucleus with $\cos(3\bar{\gamma})=0$ are indeed independent of $\sigma_{\gamma}$. The fluctuation of $\gamma$ reduces the difference between the prolate $\bar{\gamma}=0$ and the oblate $\bar{\gamma}=\pi/3$ shape. This reduction is largely described by $e^{-\frac{9 \sigma_{\gamma}^{2}}{2}}\cos (3\bar{\gamma})$, except for a small asymmetry between $\bar{\gamma}=0$ and $\bar{\gamma}=\pi/3$, clearly visible for $\lr{\left(\delta d_{\perp}/d_{\perp}\right)^3}$. 

We account for this small asymmetry by including higher-order terms in the fit function permitted by symmetry. Keeping leading and subleading terms, we have,
\begin{align}\label{eq:10}
\lr{\varepsilon_2^2 \frac{\delta d_{\perp}}{d_{\perp}}}-\lr{\varepsilon_2^2 \frac{\delta d_{\perp}}{d_{\perp}}}_{\beta=0} 
&= \left[a_0'+(a_1'\cos(3\bar{\gamma}) + b_1'\left[\sin(3\bar{\gamma})-1\right]) e^{-\frac{9 \sigma_{\gamma}^{2}}{2}}+(a_2'\left[\cos(6\bar{\gamma})+1\right] + b_2' \sin(6\bar{\gamma})) e^{-\frac{36 \sigma_{\gamma}^{2}}{2}}\right]\bar{\beta}^3\\\label{eq:11}
&= a_0+(a_1\cos(3\bar{\gamma}) + b_1\left[\sin(3\bar{\gamma})-1\right]) e^{-\frac{9 \sigma_{\gamma}^{2}}{2}}+(a_2\left[\cos(6\bar{\gamma})+1\right] + b_2\sin(6\bar{\gamma})) e^{-\frac{36 \sigma_{\gamma}^{2}}{2}}\;.
\end{align}
The same fit function is also used to describe $\lr{\left(\delta d_{\perp}/d_{\perp}\right)^3}$. The parameters in the first line and those in the second line differ by a scale factor $\bar{\beta}^3=0.28^3=0.021$. From the values of parameters displayed in Fig.~\ref{fig:1}, we concluded that the magnitude of the high-order order terms is less than 2\% of the magnitude of $a_1$ for $\lr{\varepsilon_2^2 \frac{\delta d_{\perp}}{d_{\perp}}}$ but reaches up to 5\% for $\lr{\left(\delta d_{\perp}/d_{\perp}\right)^3}$. 

Figure~\ref{fig:1} shows that the signature of triaxiality in heavy ion collisions is greatly reduced for large value of $\sigma_{\gamma}$, often found in $\gamma$-soft nuclei. A twenty-degree fluctuation in triaxiality, for example, reduces the signal by nearly 40\%. It would be difficult to distinguish between static rigid triaxial nuclei and nuclei with large fluctuations around $\bar{\gamma}=\pi/6$ using heavy ion collisions. In particular, nuclei that fluctuate uniformly between prolate and oblate shapes would give the same three-particle correlation signal as rigid triaxial nuclei! Such strong smearing also degrades the prospects of using higher-order cumulants of $\varepsilon_2$ to infer the value of $\sigma_{\gamma}$.

\begin{figure}[!h]
\begin{center}
\includegraphics[width=1\linewidth]{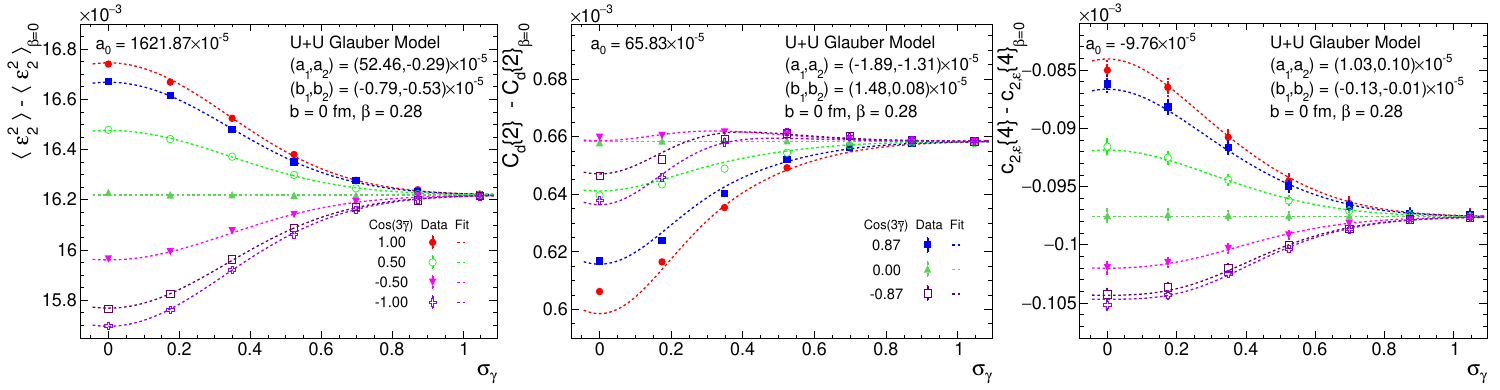}
\end{center}
\caption{\label{fig:2} The dependence of $\lr{\varepsilon_2^2}$ (left), $\lr{\left(\delta d_{\perp}/d_{\perp}\right)^2}$ (middle), and $\lr{\varepsilon_2^4}-2\lr{\varepsilon_2^2}^2$ (right) on $\sigma_{\gamma}$ for different values of $\bar{\gamma}$. The dashed lines indicate a simultaneous fit to Eq.~\eqref{eq:11}, with fit results being displayed on the plot.}
\end{figure}

For the other three observables, $\lr{\varepsilon_2^2}$, $\lr{\left(\delta d_{\perp}/d_{\perp}\right)^2}$ and $\lr{\varepsilon_2^4}-2\lr{\varepsilon_2^2}^2$, $\gamma$ dependence is known to be very weak~\cite{Jia:2021tzt}. Nevertheless, up to a few percent dependence is observed, which can also be parameterized by Eq.~\eqref{eq:10}, except that we should change $\bar{\beta}^3$ to $\bar{\beta}^2$ for the variances and to $\bar{\beta}^4$ for $\lr{\varepsilon_2^4}-2\lr{\varepsilon_2^2}^2$. However, since $\bar{\beta}$ is fixed at 0.28, all these observables can be parameterized by Eq.~\eqref{eq:11}. The data and the results of the fits are shown in Fig.~\ref{fig:2}. First, we observe that the parameter $a_0$, representing the baseline contribution associated with $\bar{\beta}$ is by far the largest, and the other terms only cause a few percent of modulation. Secondly, while the $\lr{\varepsilon_2^2}$ and $\lr{\varepsilon_2^4}-2\lr{\varepsilon_2^2}^2$ can be largely described by including the $\cos(3\bar{\gamma})$ term, the description of $\lr{\left(\delta d_{\perp}/d_{\perp}\right)^2}$ requires the inclusion of $\sin(3\bar{\gamma})$, $\cos(6\bar{\gamma})$ and $\sin(6\bar{\gamma})$ terms with comparable magnitudes. Lastly, all three observables have no sensitivity to $\bar{\gamma}$ at large $\sigma_{\gamma}$. 

\section{Impact of fluctuations in the magnitude of quadrupole deformation}

\begin{figure}[!t]
\begin{center}
\includegraphics[width=0.9\linewidth]{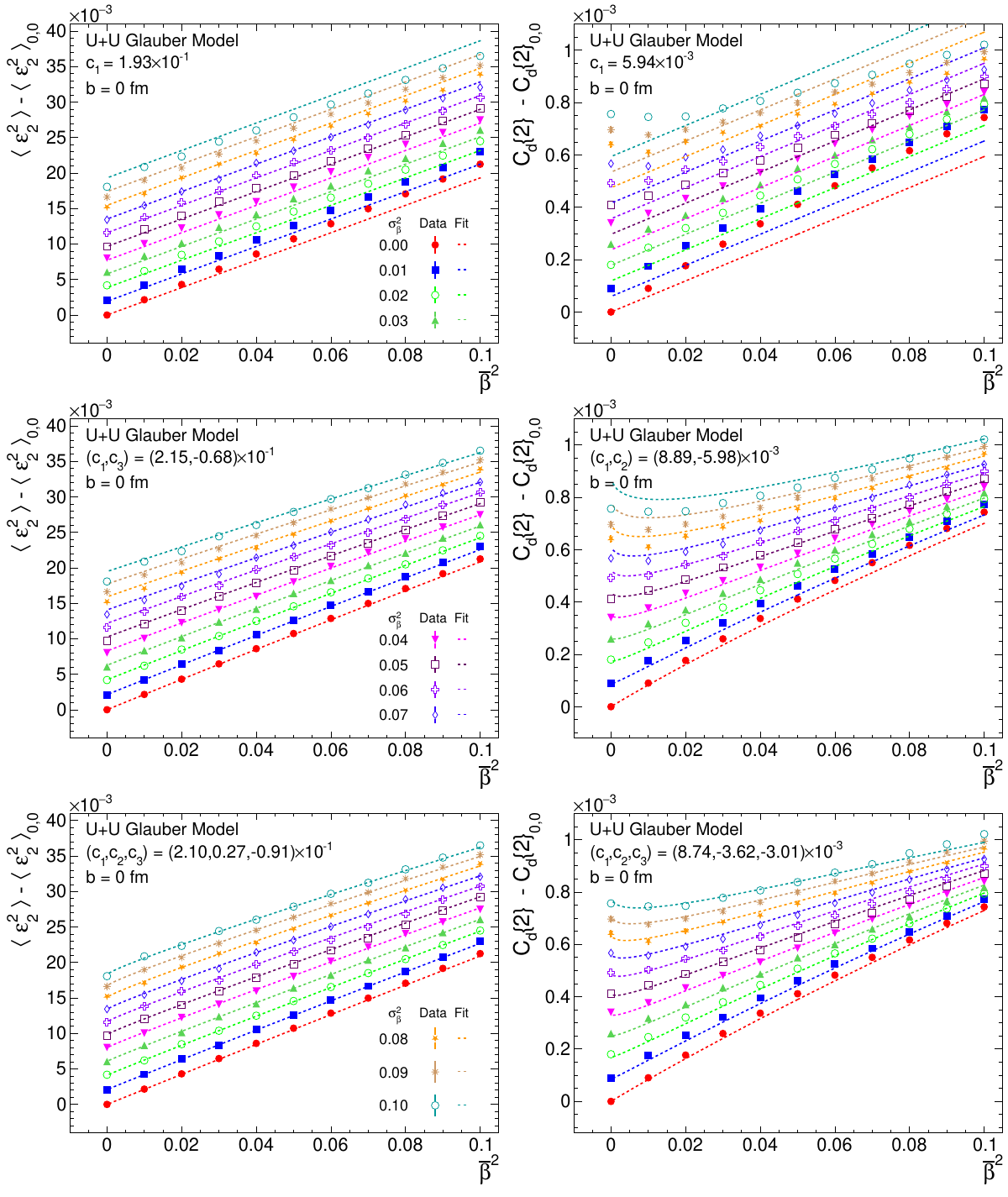}
\end{center}
\caption{\label{fig:3}  The simultaneous fit of the $\lr{\varepsilon_2^2}(\bar{\beta}, \sigma_{\beta})$ (left column) and $\lr{\left(\delta d_{\perp}/d_{\perp}\right)^2}(\bar{\beta}, \sigma_{\beta})$ (right column) calculated in U+U collisions with zero impact parameter. The top row shows the fits to Eq.~\eqref{eq:12} with only the leading term and the last row shows the fits with all three terms. The middle row show the fits including $c_1$ and $c_3$ terms for $\lr{\varepsilon_2^2}$ and $c_1$ and $c_2$ terms for $\lr{\left(\delta d_{\perp}/d_{\perp}\right)^2}$.}
\end{figure}

Next, we consider the impact of $\beta$ fluctuations. For this purpose, we shall fix the $\gamma$ to be prolate shape, e.g $\cos(3\gamma)=1$. Figure~\ref{fig:3} displays the finding for two-particle observables $\lr{\varepsilon_2^2}$ and $\lr{\left(\delta d_{\perp}/d_{\perp}\right)^2}$, again corrected by the undeformed baseline. Although approximately-linear dependencies on $\bar{\beta}^2$ are observed for both observables, the slopes of the data points also vary with $\sigma_{\beta}$. To describe this feature, we include two higher-order terms,
\begin{align}\nonumber
\lr{\varepsilon_2^2}-\lr{\varepsilon_{2}^2}_{\beta=0}\;\mathrm{or}\; \lr{\left(\frac{\delta d_{\perp}}{d_{\perp}}\right)^2}-\lr{\left(\frac{\delta d_{\perp}}{d_{\perp}}\right)^2}_{\beta=0} &= c_1\lr{\beta^2}+c_2\lr{\beta^3}+c_3\lr{\beta^4}\\\label{eq:12}
&= c_1(\bar{\beta}^2+\sigma_{\beta}^2)+c_2\bar{\beta}(\bar{\beta}^2+3\sigma_{\beta}^2)+c_3(\bar{\beta}^4+6\bar{\beta}^2\sigma_{\beta}^2+3\sigma_{\beta}^4)
\end{align}
The fits including only the leading term and all three terms are shown in the first row and the last row of Fig.~\ref{fig:3}, respectively. The fits in the middle row include the $c_1$ and $c_3$ terms for $\lr{\varepsilon_2^2}$, while they include $c_1$ and $c_2$ terms for $\lr{\left(\delta d_{\perp}/d_{\perp}\right)^2}$. Clearly, the behavior of $\lr{\left(\delta d_{\perp}/d_{\perp}\right)^2}$ at large $\bar{\beta}$ or $\sigma_{\beta}$ requires the presence of the $\lr{\beta^3}$ term in Eq.~\eqref{eq:12} with a negative coefficient $c_2<0$. In general, a large fluctuation $\sigma_{\beta}$ tends to reduce the slope of the dependence on $\bar{\beta}^2$.

\begin{figure}[!h]
\begin{center}
\includegraphics[width=0.8\linewidth]{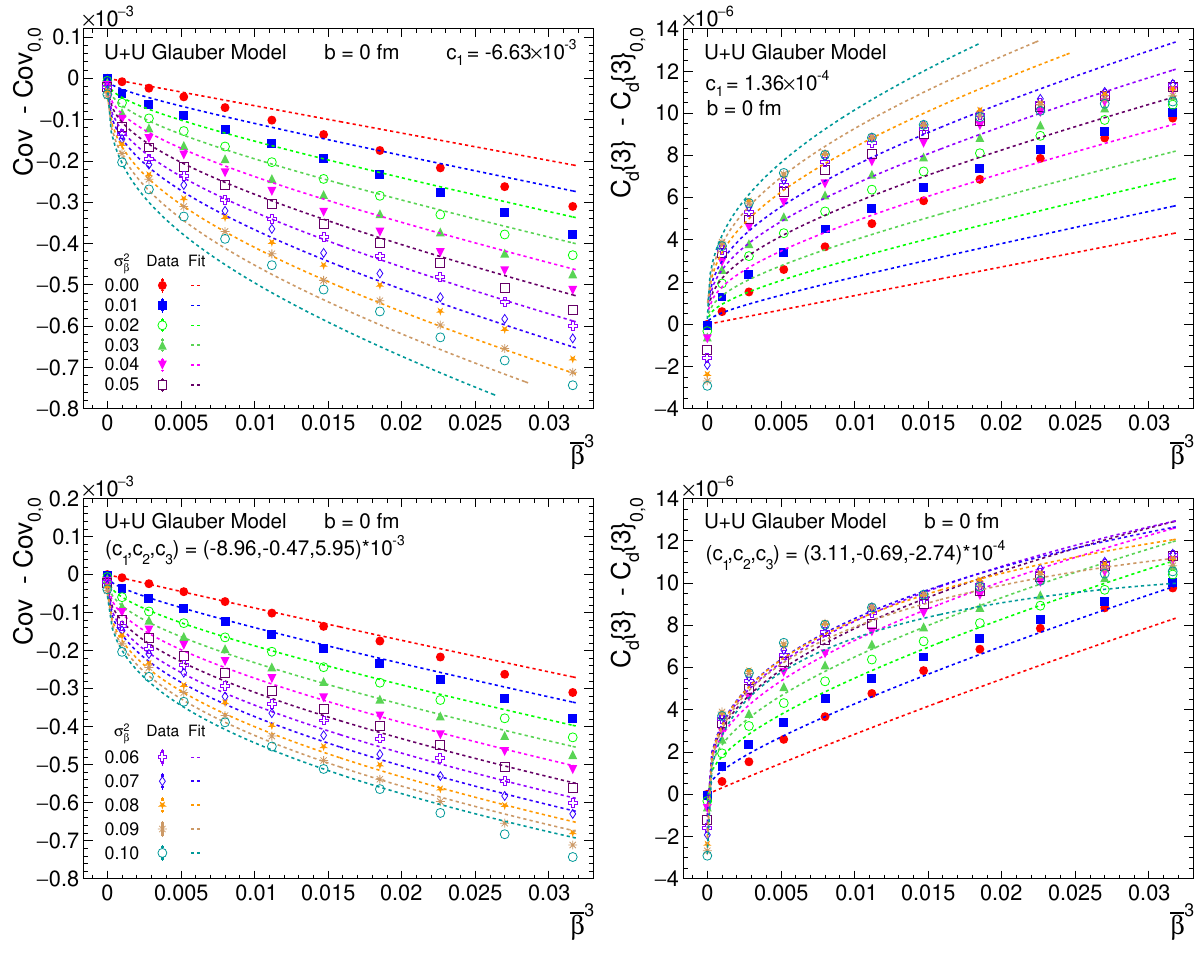}
\end{center}
\caption{\label{fig:4} The simultaneous fit of the $\lr{\varepsilon_2^2\frac{\delta d_{\perp}}{d_{\perp}}}(\bar{\beta}, \sigma_{\beta})$ (left column) and $\lr{\left(\delta d_{\perp}/d_{\perp}\right)^3}(\bar{\beta}, \sigma_{\beta})$ (right column) calculated in U+U collisions with zero impact parameter. The top row shows the results of the fit to Eq.~\eqref{eq:13} with only the leading term and the second row shows the fits with all three terms. The fit results imply that the contribution from $\lr{\beta^4}$ is negligible, though.}
\end{figure}

For the three-particle correlators, we include three terms in the fitting function as
\begin{align}\nonumber
\lr{\varepsilon_2^2\frac{\delta d_{\perp}}{d_{\perp}}}-\lr{\varepsilon_{2}^2\frac{\delta d_{\perp}}{d_{\perp}}}_{\beta=0}\;\mathrm{or}\;& \lr{\left(\frac{\delta d_{\perp}}{d_{\perp}}\right)^3}-\lr{\left(\frac{\delta d_{\perp}}{d_{\perp}}\right)^3}_{\beta=0} = c_1\lr{\beta^3\cos(3\gamma)}+c_2\lr{\beta^4\cos(3\gamma)}+c_3\lr{\beta^5\cos(3\gamma)}\\\label{eq:13}
&= [c_1\bar{\beta}(\bar{\beta}^2+3\sigma_{\beta}^2)+c_2(\bar{\beta}^4+6\bar{\beta}^2\sigma_{\beta}^2+3\sigma_{\beta}^4)+c_3(\bar{\beta}^5+10\bar{\beta}^3\sigma_{\beta}^2+15\bar{\beta}\sigma_{\beta}^4)]\cos(3\gamma)
\end{align}
The fitting results are shown in Fig.~\ref{fig:4} as a function of $\bar{\beta}^3$ for the prolate case $\cos(3\gamma)=1$. The inclusion of the high-order terms, reflecting mostly the contribution from the $\lr{\beta^5}$ component, improves the description of $\lr{\varepsilon_2^2 \frac{\delta d_{\perp}}{d_{\perp}}}$ in the region of large $\sigma_{\beta}$. However, they are not sufficient to describe the $\lr{\left(\delta d_{\perp}/d_{\perp}\right)^3}$ in the region of large $\bar{\beta}$ and $\sigma_{\beta}$. In particular, the fit also misses most data points at $\bar{\beta}=0$.  We checked that the fit can be systematically improved by including more higher moment terms, albeit only very slowly. 

Lastly, we consider the four-particle observable $c_{2,\varepsilon}\{4\}=\lr{\varepsilon_2^4}-2\lr{\varepsilon_2^2}^2$. According to findings in Fig.~\ref{fig:3}, the Taylor expansion of $\lr{\varepsilon_2^2}$ should give the first two terms as  $c_1\lr{\beta^2}+c_2\lr{\beta^4}$. Similarly, the first few terms of $\lr{\varepsilon_2^4}$ has the form of $a_1\lr{\beta^4}+a_2\lr{\beta^6}+a_3\lr{\beta^2}^2+a_4\lr{\beta^2}\lr{\beta^4}$. Therefore, the natural expression for $c_{2,\varepsilon}\{4\}$ up to second order correction should be
\begin{align}\nonumber
&c_{2,\varepsilon}\{4\}-c_{2,\varepsilon}\{4\}_{\beta=0} = a_1\lr{\beta^4}+a_2\lr{\beta^6}+a_3\lr{\beta^2}^2 +a_4\lr{\beta^2}\lr{\beta^4}-(c_1\lr{\beta^2}+c_2\lr{\beta^4})^2\approx  a_1\lr{\beta^4}-b_1\lr{\beta^2}^2 
+a_2\lr{\beta^6} -b_2\lr{\beta^2}\lr{\beta^4}\\\label{eq:14}
&=a_1 (\bar{\beta}^4+6\bar{\beta}^2\sigma_{\beta}^2+3\sigma_{\beta}^4) - b_1 (\bar{\beta}^2+\sigma_{\beta}^2)^2 + a_2(\bar{\beta}^6+15\bar{\beta}^4\sigma_{\beta}^2+45\bar{\beta}^2\sigma_{\beta}^4 +15 \sigma_{\beta}^6) - b_2(\bar{\beta}^2+\sigma_{\beta}^2)(\bar{\beta}^4+6\bar{\beta}^2\sigma_{\beta}^2+3\sigma_{\beta}^4)
\end{align}
with $b_1 = c_1^2-a_3$ and $b_2 = 2c_1c_2-a_4$. The leading order correction includes the first two terms with $a_1$ and $b_1$, while the remaining two terms are the subleading-order corrections.

\begin{figure}[!h]
\begin{center}
\includegraphics[width=0.8\linewidth]{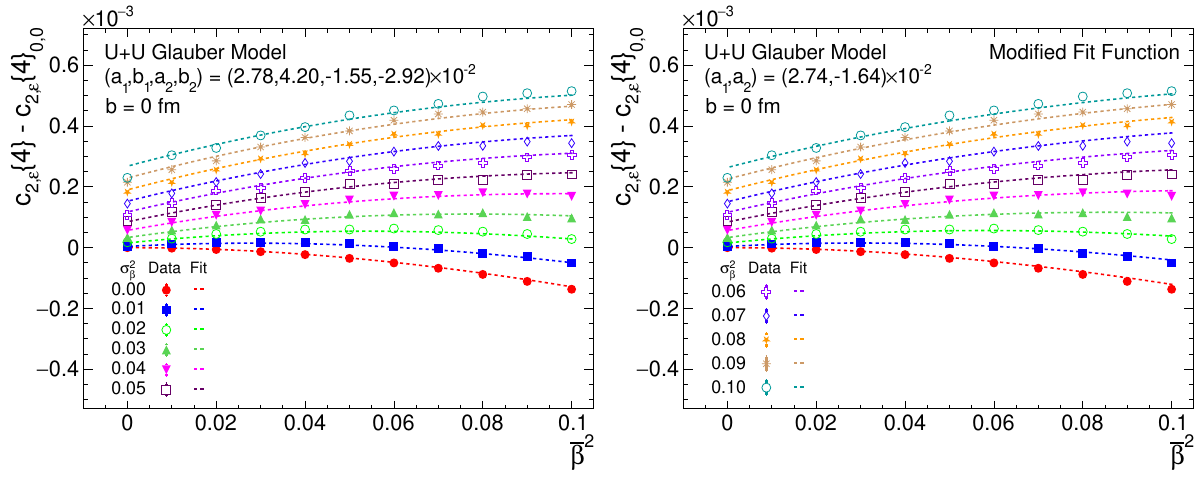}
\end{center}
\caption{\label{fig:5} The fit of the $c_{2,\varepsilon}\{4\}(\bar{\beta}, \sigma_{\beta})$ data calculated in U+U collisions with zero impact parameter to Eq.~\eqref{eq:14} (left) and Eq.~\eqref{eq:15} (right).}
\end{figure}

The results from the Glauber model and the fit to Eq.~\eqref{eq:14} are shown in the left panel of Fig.~\ref{fig:5}. The strong variation of $c_{2,\varepsilon}\{4\}$ with both $\bar{\beta}$ and $\sigma_{\beta}$ is captured nicely by the fit. For small values of $\sigma_{\beta}$, the deformation has a negative contribution to $c_{2,\varepsilon}\{4\}$ that is proportional to $\bar{\beta}^4$. Even for a relatively small $\sigma_{\beta}$ value, $c_{2,\varepsilon}\{4\}$ becomes positive. A previous study shows that the centrality fluctuation also tends to give a positive value of $c_{2,\varepsilon}\{4\}$~\cite{Zhou:2018fxx}. Therefore, a negative $c_{2,\varepsilon}\{4\}$ which decreases further in central collisions would be an unambiguous indication for a large static quadrupole deformation of the colliding nuclei. 

The values of the fit parameters show some interesting relations, i.e. $b_1\approx 3a_1/2$ and $b_2 \approx 2a_2$. This means that the distribution can also be described by the following alternative form,
\begin{align}\label{eq:15}
c_{2,\varepsilon}\{4\}-c_{2,\varepsilon}\{4\}_{\beta=0} &\approx \frac{a_1}{2} (6\bar{\beta}^2\sigma_{\beta}^2+3\sigma_{\beta}^4-\bar{\beta}^4) +a_2 (\bar{\beta}^4\sigma_{\beta}^2+27\bar{\beta}^2\sigma_{\beta}^4 +9 \sigma_{\beta}^6-\bar{\beta}^6)
\end{align}
The contribution of residual terms is only a few percent. Indeed, a fit of this form describes the data very well as shown in the right panel of Fig.~\ref{fig:5}. This behavior provides a clear intuition on how the fluctuation terms containing $\sigma_{\beta}$ compete with the terms containing only $\bar{\beta}$. For example, assuming $\bar{\beta}=\sigma_{\beta}$, the contribution from fluctuation-related terms is a factor of 9 (37) times the $\bar{\beta}^4$ ($\bar{\beta}^6$) in the leading-order (subleading order). Thus, even a relatively small fluctuation could have a strong impact on $c_{2,\varepsilon}\{4\}$. Note that the liquid-drop model results in Table~\ref{tab:1} predict $b_1 = 7a_1/5$, slightly smaller than the Glauber model expectation.

Experimentally, we can measure $\lr{v_2^2}$ and $\lr{v_4^2}$, which are linearly related to $\lr{\varepsilon_{2}^2}$ and $\lr{\varepsilon_{2}^4}$, respectively. Thus, it is natural to ask whether one could constrain the $\bar{\beta}$ and $\sigma_{\beta}$ from these two quantities. So far we have learned that the combination in the cumulant definition $c_{2,\varepsilon}\{4\}=\lr{\varepsilon_2^4}-2\lr{\varepsilon_2^2}^2$ is not sufficient to achieve such separation. Motivated by this fact, we tried a more general combination $f(\bar{\beta},\sigma_{\beta};k)=\lr{\varepsilon_2^4}-k\lr{\varepsilon_2^2}^2$, and identify the $k$ value for which the $f(\bar{\beta},\sigma_{\beta};k)$ have the least variation in $\sigma_{\beta}$. The best value found is $k=k_{0}=2.541$, for which the data points follow an approximately-linear dependence $\bar{\beta}^4$ as shown in Fig.~\ref{fig:6}. A similar study using the quark Glauber model gives a nearly identical $k_0$ value (see appendix). The data points yet do not fully collapse on a single curve, implying a small remaining $\sigma_{\beta}$ dependence. The amount of spread is estimated to be about 25\% relative for a given $\bar{\beta}$, corresponding to a variation of $\bar{\beta}$ of about $1-\sqrt[4]{0.75}=7$\%. This 7\% value is the best precision for determining $\bar{\beta}$ in the Glauber model using this method.  The determined $\bar{\beta}$ value can then be plugged into Eq.~\eqref{eq:12} (considering only the leading order is sufficient for $\lr{\varepsilon_2^2}$ as shown in Fig.~\ref{fig:3}) to determine $\sigma_{\beta}$.

\begin{figure}[!t]
\begin{center}
\includegraphics[width=0.5\linewidth]{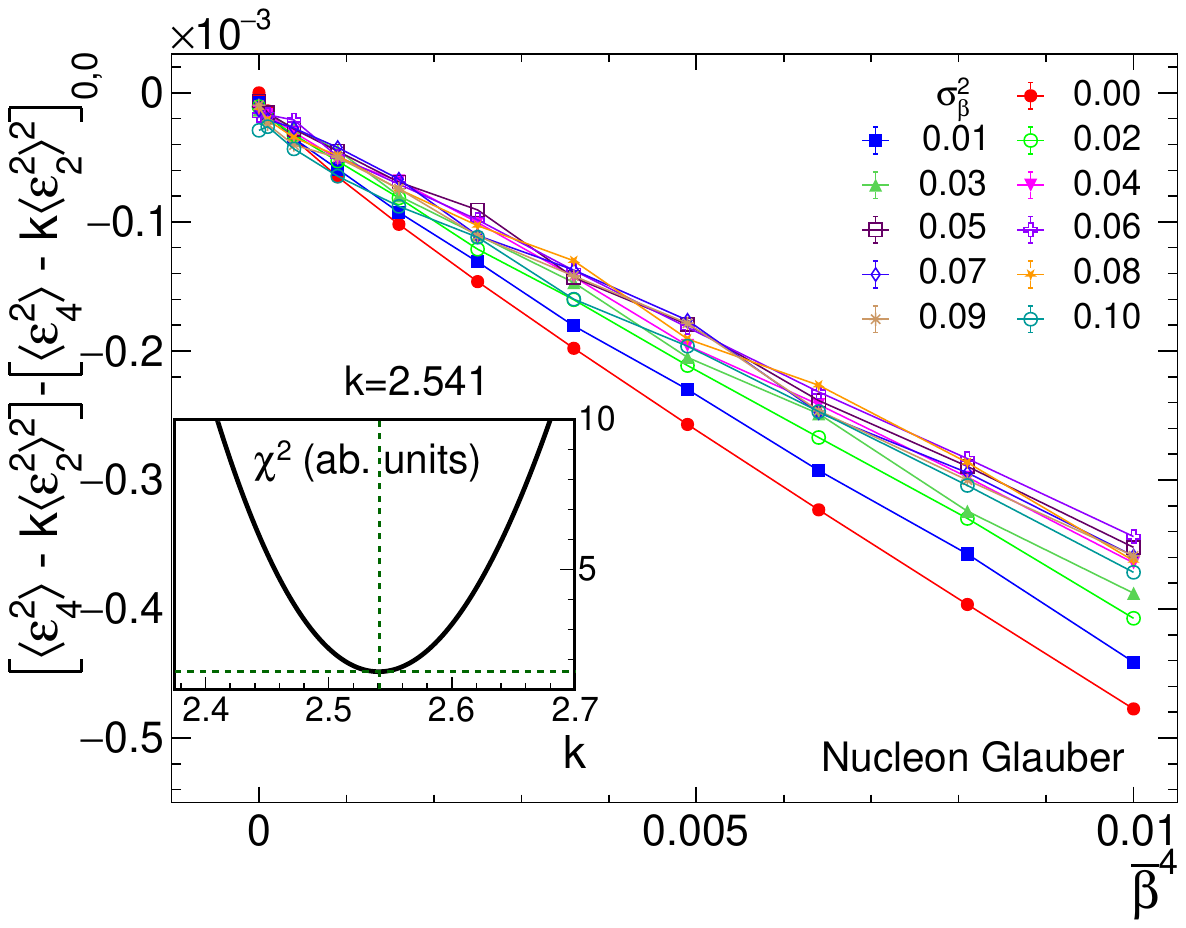}
\end{center}
\caption{\label{fig:6} The values of $\lr{\varepsilon_2^4}-K\lr{\varepsilon_2^2}^2$ for the value of $K$ that minimize the dependence on $\sigma_{\beta}$ in the nucleon Glauber model. The insert panel shows the $K$ dependence of $\chi^2$, which is calculated as $\chi^2 = \sum_{i} \sum_j (f_{ij}-\bar{f}_{i})^2/\sigma_{i.j}^2$, where $f_{ij}=f(\bar{\beta}_i,\sigma_{\beta,j}$, $\bar{f}_{i} =  \sum_j f_{ij}/\sum_j$ and $\sigma_{ij}$ is the statistical error bar on the $i,j$-th data point.}
\end{figure}

\section{Impact of fluctuations on normalized quantities}

In the study of flow fluctuations in heavy ion collisions, it is often desirable to calculate the normalized quantities between high-order cumulant and low-order cumulants, which have the advantage of canceling the final state effects. Here we study three quantities following the convention from Ref.~\cite{Jia:2021qyu}, which is different from that in Ref.~\cite{Bozek:2016yoj}),
\small
\begin{align}\label{eq:16}
\rho = \frac{\lr{\varepsilon_2^2\frac{\delta d_{\perp}}{d_{\perp}}}-\lr{\varepsilon_{2}^2\frac{\delta d_{\perp}}{d_{\perp}}}_{\beta=0}}{\left(\lr{\varepsilon_2^2}-\lr{\varepsilon_{2}^2}_{\beta=0}\right) \sqrt{\lr{\left(\frac{\delta d_{\perp}}{d_{\perp}}\right)^2}-\lr{\left(\frac{\delta d_{\perp}}{d_{\perp}}\right)^2}_{\beta=0}}}\;, \mbox{nc}_{d}\{3\} = \frac{\lr{(\frac{\delta d_{\perp}}{d_{\perp}})^3}-\lr{(\frac{\delta d_{\perp}}{d_{\perp}})^3}_{\beta=0}}{\left(\lr{\left(\frac{\delta d_{\perp}}{d_{\perp}}\right)^2}-\lr{\left(\frac{\delta d_{\perp}}{d_{\perp}}\right)^2}_{\beta=0}\right)^{3/2}}\;, \mbox{nc}_{\varepsilon}\{4\} = \frac{c_{2,\varepsilon}\{4\}-c_{2,\varepsilon}\{4\}_{\beta=0}}{\left(\lr{\varepsilon_2^2}-\lr{\varepsilon_{2}^2}_{\beta=0}\right)^2}
\end{align}\normalsize
Since $^{96}$Zr has little quadruple deformation $\beta_{\rm Zr}\approx0$, these quantities can be constructed directly from measurements in $^{96}$Ru+$^{96}$Ru and $^{96}$Zr+$^{96}$Zr collisions.

\begin{figure}[!t]
\begin{center}
\includegraphics[width=1\linewidth]{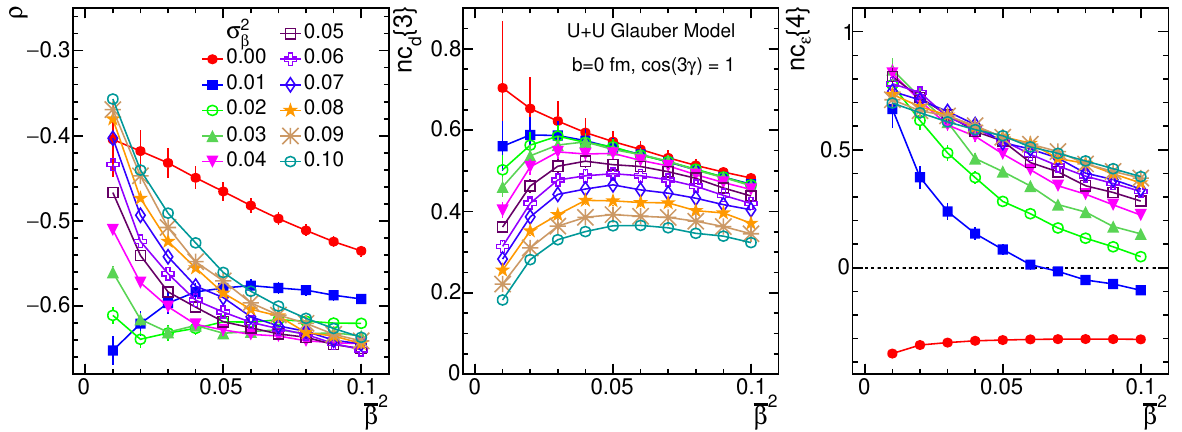}
\end{center}
\caption{\label{fig:7} The normalized three-particle correlators, $\rho$ (left) and ${\mathrm nc}_d\{3\}$ (middle) and normalized four-particle correlator ${\mathrm nc}_{\varepsilon}\{4\}$ (right) defined in Eq.~\eqref{eq:12} as a function of $\bar{\beta}^2$ for different values of $\sigma_{\beta}^2$.}
\end{figure}

The impact of $\beta$ fluctuation is shown in Fig.~\ref{fig:7} for prolate nuclei $\cos(3\gamma)=1$. For sufficiently large values of $\bar{\beta}$, the correlator $\rho$ becomes nearly independent of $\bar{\beta}$ and have a weak dependence on $\sigma_{\beta}$. In the large $\bar{\beta}$ region, $\rho$ quickly converges to a value around $-0.62$ nearly independent of $\sigma_{\beta}$. In the moderate $\bar{\beta}$ region say $\bar{\beta}\sim0.2$, the $\rho$ first decreases quickly to a value around $-0.6$, but then increases gradually with $\sigma_{\beta}$. The values of ${\mathrm nc}_d\{3\}$ have similar convergence trends towards large $\bar{\beta}$ around 0.4, but much more slowly compare to $\rho$. The ${\mathrm nc}_{\varepsilon}\{4\}$ has a negative and nearly constant value when $\sigma_{\beta}=0$, while it increases rather quickly with $\sigma_{\beta}$. Even for a value of $\sigma_{\beta}^2=0.01$, the ${\mathrm nc}_{\varepsilon}\{4\}$ stays positive until $\bar{\beta}^2>0.06$. For larger values of $\sigma_{\beta}^2$, the ${\mathrm nc}_{\varepsilon}\{4\}$ decreases with increasing $\bar{\beta}^2$, but always remains positive over the range of $\bar{\beta}$ studied. 

\begin{figure}[!t]
\begin{center}
\includegraphics[width=1\linewidth]{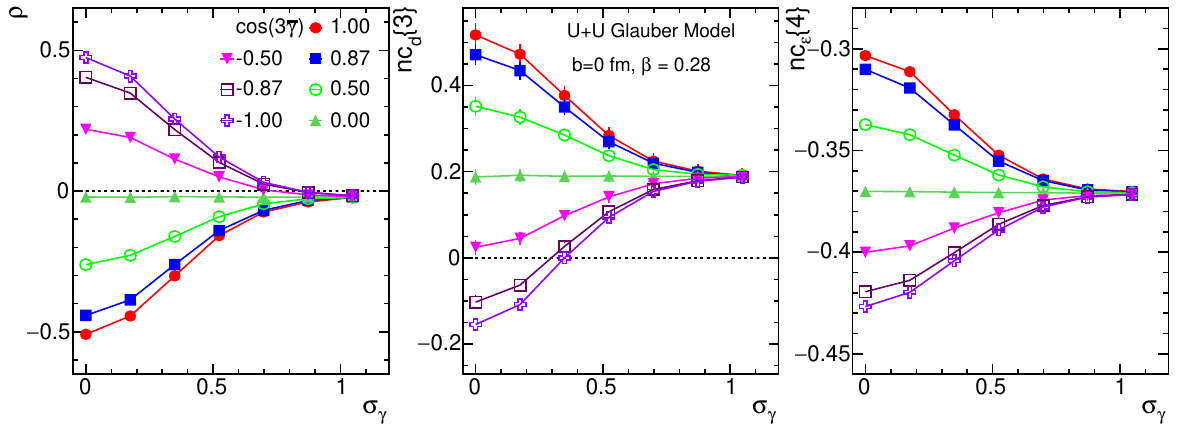}
\end{center}
\caption{\label{fig:7b} The normalized three-particle correlators, $\rho$ (left) and ${\mathrm nc}_d\{3\}$ (middle) and normalized four-particle correlator ${\mathrm nc}_{\varepsilon}\{4\}$ (right) defined in Eq.~\eqref{eq:12} as a function of $\sigma_{\gamma}$ for different values of $\cos(3\bar{\gamma})$.}
\end{figure}

Figure~\ref{fig:7b} shows the impact of $\gamma$ fluctuation calculated by assuming $\beta=0.28$. The trends of the data are very similar to those shown in Figs.~\ref{fig:1} and ~\ref{fig:2}. The values of these observables are sensitive to $\bar{\gamma}$ only when $\sigma_{\gamma}$ are not very large. Also, they also do not depends on $\sigma_{\gamma}$ when $\cos(3\bar{\gamma})=0$. 

\section{summary}

We studied the impact of fluctuations of nuclear quadrupole deformation on several heavy-ion observables in a Monte Carlo Glauber model. In particular, we focus on eccentricity $\varepsilon_2$ and inverse size $d_{\perp}$ in each event, which can be related to the event-wise elliptic flow and mean transverse momentum in the final state. The triaxiality $\gamma$ has a strong impact on three-particle correlators, but the impact diminishes for larger $\sigma_{\gamma}$. In particular, when $\sigma_{\gamma}$ is large, the observables do not distinguish between prolate deformation and oblate deformation, i.e. the values of all observables approach those obtained in collisions of rigid triaxial nuclei with the same $\beta$.  The mean and standard deviation of quadrupole fluctuations, $\bar{\beta}$ and $\sigma_{\beta}$, have a strong influence on all observables. The influence on two-particle observables $\lr{\varepsilon_2^2}$ and $\lr{(\delta d_{\perp}/d_{\perp})^2}$ is proportional to $\lr{\beta^2}=\bar{\beta}^2+\sigma_{\beta}^2$, however, the $\lr{(\delta d_{\perp}/d_{\perp})^2}$ also has a sizable subleading order term proportional to $\lr{\beta^3}$. The three-particle observables to the leading order are proportional to $\lr{\cos(3\gamma)\beta^3} = \cos(3\gamma) \bar{\beta}(\bar{\beta}+3\sigma_{\beta}^2) $, whereas the four-particle observables to the leading order are proportional to  $\lr{\beta^4} = \bar{\beta}^4+6\bar{\beta}^2\sigma_{\beta}^2+3\sigma_{\beta}^4$. Hence, the standard deviation of $\beta$ fluctuation has a stronger impact than $\bar{\beta}$ for these higher-order observables.  

By combining two and four-particle cumulant of $\varepsilon_2$, we have constructed a simple formula to constrain parameters $\bar{\beta}$ and $\sigma_{\beta}$ simultaneously. Such separation becomes less effective when $\sigma_{\beta}$ is comparable or larger than $\bar{\beta}$. In the future, it would be interesting to carry out a full hydrodynamic model simulation to quantify the efficacy of this method on the final state flow observables.

{\bf Acknowledgment:} This research is supported by DOE DE-FG02-87ER40331.

\section*{Appendix}
\begin{figure}[!h]
\begin{center}
\includegraphics[width=1\linewidth]{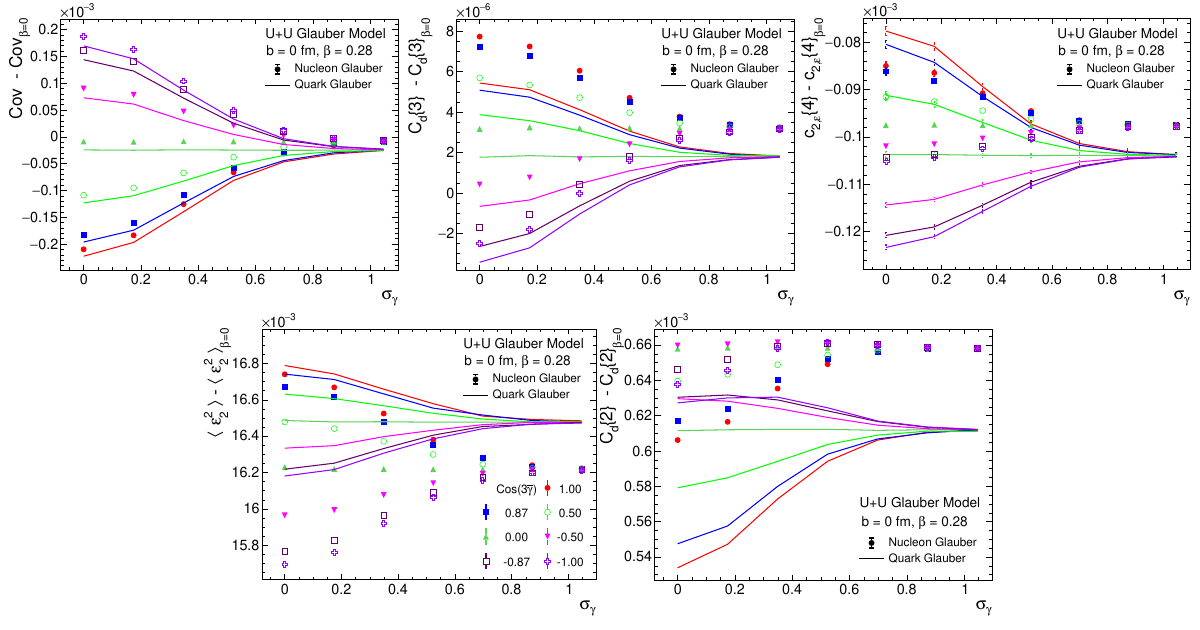}
\end{center}
\caption{\label{fig:8} Comparison of the five observables between nucleon Glauber model (symbols) and quark Glauber model (lines with matching colors) as a function of $\sigma_{\gamma}$ for different values of $\bar{\gamma}$.}
\end{figure}

\begin{figure}[!h]
\begin{center}
\includegraphics[width=1\linewidth]{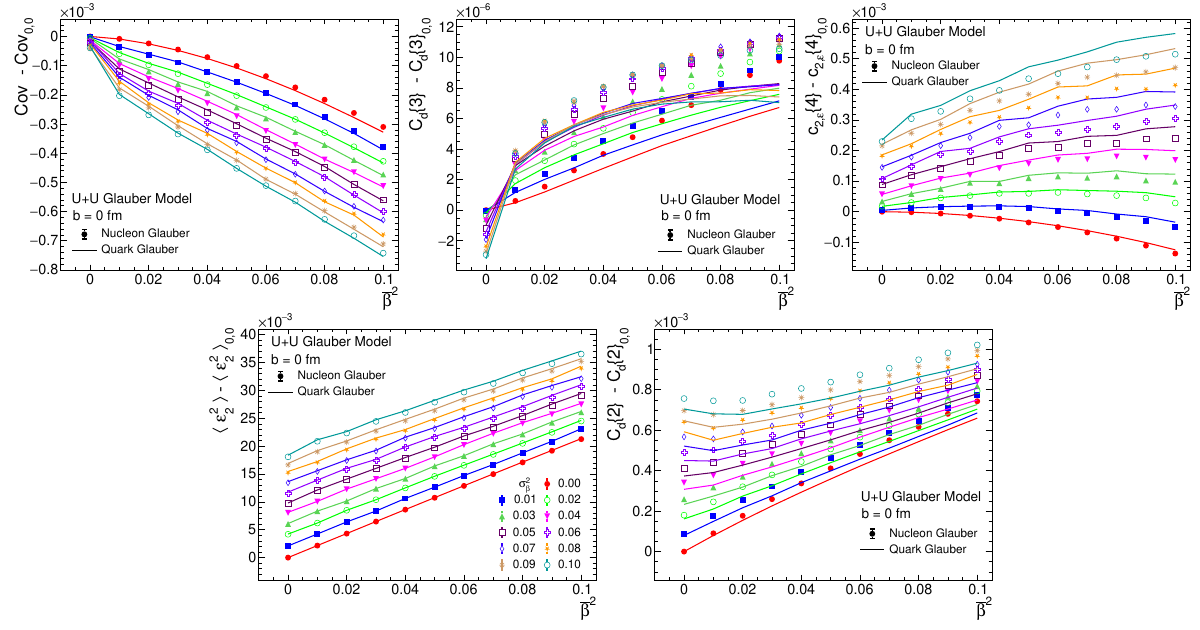}
\end{center}
\caption{\label{fig:9} Comparison of the five observables between nucleon Glauber model (symbols) and quark Glauber model (lines with matching colors) as a function of $\bar{\beta}^2$ for different values of $\sigma_{\beta}$. }
\end{figure}

The default results in this paper are obtained with the nucleon Glauber model.  We have repeated the analysis for the quark Glauber model and compared it with the nucleon Glauber model results in Figs.~\ref{fig:8} and \ref{fig:9} for the impact of $\gamma$ fluctuation and $\beta$ fluctuation, respectively. The trends are mostly very similar. A few exceptions are observed. In particular, the results of the two models are shifted vertically from each other in Fig.~\ref{fig:8}. In the case of $\beta$ fluctuation in Fig.~\ref{fig:9}, the variance $c_{\rm d}\{2\}$ and skewness $c_{\rm d}\{3\}$ are systematically different between the two models in the high $\bar{\beta}$ region. Table~\ref{tab:2} gives the cumulant expression for the case where the projectile and target nuclei have the same mass number but different deformations. These expressions are expected from the additive nature of the cumulants. 

\begin{figure}[!t]
\begin{center}
\includegraphics[width=0.5\linewidth]{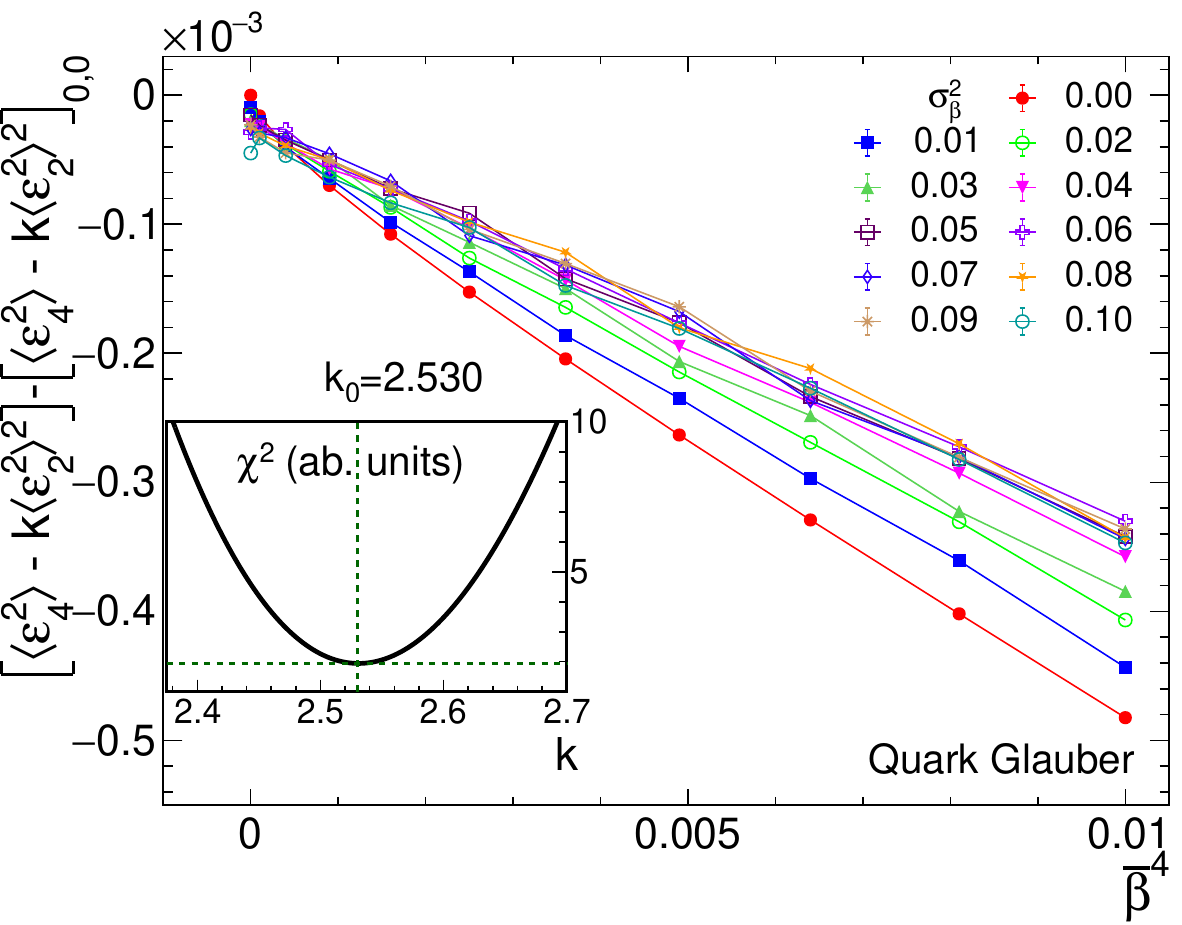}
\end{center}
\caption{\label{fig:10} The values of $\lr{\varepsilon_2^4}-K\lr{\varepsilon_2^2}^2$ for the value of $K$ that minimize the dependence on $\sigma_{\beta}$ in the quark Glauber model, similar to Fig.~\ref{fig:6}.}
\end{figure}

\begin{table}[!h]
\centering
\begin{tabular}{c||c}\hline
Cumulants& Liquid-drop model estimate\\\hline
\multirow{2}{*}{$\lr{(\delta d_{\perp}/d_{\perp})^2}$}&\multirow{2}{*}{$\frac{1}{64\pi}\left(\lr{\beta_p^2}+\lr{\beta_t^2}\right)$}\\ &\\\hline
\multirow{2}{*}{$\lr{(\delta d_{\perp}/d_{\perp})^3}$}& \multirow{2}{*}{$\frac{\sqrt{5}}{1792 \pi^{3/2}}\left(\lr{\cos(3\gamma_p)\beta_p^3}+\lr{\cos(3\gamma_t)\beta_t^3}\right)$}\\ &\\\hline
\multirow{2}{*}{$\lr{(\delta d_{\perp}/d_{\perp})^4}-3\lr{(\delta d_{\perp}/d_{\perp})^2}^2$}&\multirow{2}{*}{$-\frac{3}{28672 \pi^{2}}\left(7\lr{\beta_p^2}^2+7\lr{\beta_t^2}^2-5\lr{\beta_p^4}-5\lr{\beta_t^4}\right)$}\\ &\\\hline
\multirow{2}{*}{$\lr{\varepsilon_2^2}$} & \multirow{2}{*}{$\frac{3}{8\pi}\left(\lr{\beta_p^2}+\lr{\beta_t^2}\right)$}\\ &\\\hline
\multirow{2}{*}{$\lr{\varepsilon_2^4}-2\lr{\varepsilon_2^2}^2$} & \multirow{2}{*}{$-\frac{9}{224\pi^2}\left(7\lr{\beta_p^2}^2+7\lr{\beta_t^2}^2-5\lr{\beta_p^4}-5\lr{\beta_t^4}\right)$}\\  &\\\hline
\multirow{2}{*}{$\left(\lr{\varepsilon_2^6}-9\lr{\varepsilon_2^4}\lr{\varepsilon_2^2}+12\lr{\varepsilon_2^2}^3\right)/4$}& 
\multirow{2}{*}{$\frac{81}{512\pi^3}\left[\left(\lr{\beta^2}^3-\frac{15}{14}\lr{\beta^4}\lr{\beta^2} +\frac{1175}{6006}\lr{\beta^6}-\frac{25}{3003}\lr{\cos(6\gamma)\beta^6}\right)_p+\left(\vphantom{\lr{\beta^2}^3}cc.\right)_t\right]$}\\ &\\\hline
\multirow{2}{*}{$\lr{\varepsilon_2^2(\delta d_{\perp}/d_{\perp})}$} & \multirow{2}{*}{$-\frac{3 \sqrt{5}}{224\pi^{3/2}} \left(\lr{\cos(3\gamma_p)\beta_p^3}+\lr{\cos(3\gamma_t)\beta_t^3}\right)$}\\ &\\\hline
\multirow{2}{*}{$\lr{\varepsilon_2^2(\delta d_{\perp}/d_{\perp})^2}-\lr{\varepsilon_2^2}\lr{(\delta d_{\perp}/d_{\perp})^2}$} & \multirow{2}{*}{$-\frac{3}{3584\pi^2}\left(7\lr{\beta_p^2}^2+7\lr{\beta_t^2}^2-5\lr{\beta_p^4}-5\lr{\beta_t^4}\right)$}\\ &\\\hline
\multirow{2}{*}{$\lr{ {\bm \epsilon}_2^2{\bm \epsilon}_4^*}$} & \multirow{2}{*}{$\frac{45}{112\pi^2}\left(\lr{\beta_p^4}+\lr{\beta_t^4}\right)$}\\ &\\\hline
\end{tabular}
\caption{\label{tab:2} The leading-order results of various cumulants calculated for the nucleus with a sharp surface by assuming different deformations for the two colliding nuclei}.
\end{table}

\clearpage

\bibliography{shapefluc}{}
\bibliographystyle{apsrev4-1}
\end{document}